\documentclass[preprint,12pt]{elsarticle}
\usepackage{amsmath}
\usepackage{amssymb}
\usepackage{amscd}

\begin{document}
\title{ Qualitative features of the rearrangement 
of molecular energy spectra from a  
``wall-crossing'' perspective.}
\author[TI]{T. Iwai}
\ead{iwai@amp.i.kyoto-u.ac.jp}
\author[BZ]{ B.~Zhilinskii\corref{cor1} }
\ead{zhilin@univ-littoral.fr}
\cortext[cor1]{Corresponding author}
\address[TI]{Kyoto University, Kyoto, Japan }
\address[BZ]{Universit\'e du Littoral C\^ote d'Opale, 59140 Dunkerque, France} 

\begin{abstract}
 Qualitatively  different systems of molecular energy bands are  studied on 
example of a parametric family of effective Hamiltonians
 describing rotational structure of triply degenerate vibrational state of
 a cubic symmetry molecule. The modification of band structure under 
 variation of control parameters is associated with a topological invariant 
 ``delta-Chern". 
This invariant is evaluated by using a local Hamiltonian for the control 
parameter values assigned at the boundary between adjacent parameter domains 
which correspond to qualitatively different band structures. 
\end{abstract}
\maketitle

\section{Introduction}
The purpose of the present article is to discuss qualitative models describing 
small quantum  systems of a finite number of particles 
and their modifications under a variation of control parameters
on the basis of  effective phenomenological Hamiltonians. 
The qualitative analysis of models along a change in parameters 
is based on the classical-quantum correspondence 
and it is to be stressed that the transition from one
qualitative regime to another is similar, in the sense of basic idea, to much more general mathematical  
``wall crossing" formalism \cite{KontsSob,Gaiotto} 
developed  essentially in relation to  
physical models suggested in string theory.    

The general idea of qualitative description
of simple quantum systems consists in studying a family of objects (or models) 
depending on a number of control parameters  (external ones like characteristics of
external potentials or  internal ones characterizing through approximate dynamical
parameters the  effective behavior of the model). 

A realization  of the qualitative approach consists in finding characteristics which are 
defined almost everywhere ({\it i.e.}, for almost all values of control parameters)
and are piece-wise constant on the space of control parameters.  
In other words, the qualitative description assumes the splitting of the space of control 
parameters into disjoint regions by a codimension one boundary and assumes equally
the existence of boundaries of higher codimension. 
One of such qualitative characteristics of molecular systems to be 
studied in the present paper is the energy band 
structure which exists for some  regions in the control parameter space and 
changes discontinuously when crossing the boundary of the region 
\cite{VPVdp,PhysRev93,FaurePRL,IwaiAnnPhys}. 
Another example of a qualitative characteristic  
is the number of non-linear normal modes \cite{StewartRM,CP2Zhil}, or the number of 
stationary points of an effective Hamiltonian 
in the classical limit, whose modification is associated with
quantum bifurcations \cite{QuantBif}. 

\begin{figure}
\begin{center}
\includegraphics[width=0.5\columnwidth]{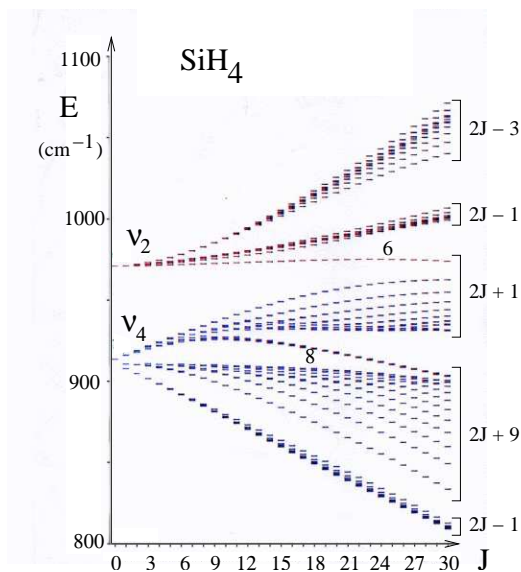} 
\end{center}
\caption{Reduced rovibrational energy as a function of rotational quantum number $J$ 
for vibrational modes
$\nu_4$ ($F_2$ symmetry type) and $\nu_2$ ($E$ symmetry type) 
of SiH$_4$ tetrahedral molecule ($T_d$ point symmetry group). 
The numbers of energy levels within clearly seen energy bands 
are indicated for $J\sim30$ region. The numbers  
of energy levels (6 and 8) in the rotational clusters redistributed
between the bands are also shown.
\label{FigEx3States} }
\end{figure}

The concept of energy bands is most frequently associated  with solid state
problems and with the existence of periodic symmetry and gaps in the distribution of 
energy states \cite{Kitaev,Wen}.  The present paper deals mainly with 
finite particle systems which exhibit the
presence of energy bands due to the splitting of relevant dynamical  
variables into two qualitatively different types in association with low- and high-energy
excitations \cite{FaureLMP}. 
On using the well known Born-Oppenheimer (or adiabatic) approximation, 
this splitting can be classically interpreted as a splitting into ``slow" and
``fast" variables. ``Slow" variables describe internal structure of bands
associated with low-energy excitations, while ``fast" variables are related 
to high-energy excitations and correspond to passage from one band to another, {\it i.e.}, 
they are responsible for inter-band structure.  To simplify the analysis and discussions,  
we suppose that in the partial classical limit with respect to the slow variables 
the quantum system under study is described on the compact classical phase space and consequently 
the resultant semi-quantum system possesses discrete quantum spectrum 
consisting of a finite number of energy levels distributed among a finite number of bands.  
In this simple situation,  the modification of the band structure can be associated with 
the redistribution of energy levels between bands and consequently with the modification of the 
number of energy levels within bands. 
As an example of such phenomenon, Figure \ref{FigEx3States} 
reproduces the rotational structure of two vibrational states
(doubly degenerate $\nu_2$ state of $E$ symmetry and triply degenerate $\nu_4$
state of $F_2$ symmetry) of the tetrahedral molecule SiH$_4$, 
reconstructed from high resolution
infrared spectroscopic data \cite{Dijon}. 
The rotational quantum number, $J$,
serves as a control parameter.
Two rearrangements of energy levels between bands are seen on this diagram. 
As $J$ increases, one group of eight energy levels goes from 
the upper branch of $\nu_4$ mode to
the middle branch, and at slightly higher $J$ values another
group of six energy levels passes from the lower branch of $\nu_2$ mode to
the upper branch of $\nu_4$ mode. 
Although the behavior of energy levels
in Fig. \ref{FigEx3States} is gradual and do not show discontinuities,
the transformation of the model system into a semi-quantum system through
classical limit for rotational variables and the
introduction of vector bundle description allow us to find an integer
topological invariant, the Chern number for eigen-line bundle which exibits
piece-wise constant behavior as a function of $J$ and characterizes
the qualitative modification of band structure.

Many similar phenomena are well-known in molecular physics 
\cite{MolPhys88,PhysRev93,PhysRep2}  and 
a number of high-precision experimental data are available.
The example we treat as a model below  is quite close to the model \cite{Dhont2000} which describes 
the rearrangement of rotation-vibration bands in Mo(CO)$_6$ molecule 
on the basis of the extensive high resolution experimental
data from \cite{Exp2000}.  At the same time,  a general qualitative understanding and 
a mathematical description of these and similar more general phenomena such as the fusion of several 
bands into one  (see \cite{FaureCP2}) or the inverse splitting of one band into several individual 
ones  are not yet properly formulated. 

\section{Conceptual setting up and a way of analysis} 
Recent works on quantum-classical correspondence for finite particle systems
exhibiting a band structure (rotational structure of several vibrational states of an
isolated molecule is a typical example of such a structure) are based on the procedure 
of the aforementioned  partial classical limit for a particular quantum system so that the resultant 
semi-quantum system can be described in terms of vector bundle.   
The  base space of this bundle is associated with classical variables describing
internal structure of bands, whereas fibers are associated with quantum states
forming different bands \cite{IwaiHolonomy,FaurePRL,FaureAAM,IwaiAnnPhys}.

We restrict our study in this article to vector bundles whose base space is a 
compact classical phase space of one degree of freedom. 
As a natural molecular example, the real two-dimensional sphere $S^2$ can serve as 
the phase space associated with molecular rotation. 
An Hermitian matrix is used to treat the complete problem in a semi-quantum model 
 \cite{SadZhilMonodr,PhysRep2}.  
Its eigenvectors assigned to all the points of $S^2$ form generically isolated line-bundles, 
which we call eigen-line bundles.  
This is because typically a Hermitian matrix has nondegenerate eigenvalues,  
since the codimension of degeneracies for a parametric family of Hermitian matrices is three 
(real) \cite{ArnoldChern} whereas the dimension of the base space 
in the considered case is only 2 (real).

Consequently, an isolated semi-quantum model Hamiltonian has  non-degenerate eigenvalues 
everywhere on $S^2$, and the trivial vector bundle over $S^2$ can split into a direct sum of  
eigen-line bundles describing individual isolated bands. 
The topological invariant characterizing the qualitative structure
of bands for such model is the system of the first Chern numbers for the set of eigen-line
bundles. The relation between Chern numbers of eigen-line bundles for a semi-quantum
model and the numbers of
quantum levels within bands for parent purely quantum problem was demonstrated 
in \cite{FaurePRL,FaureLMP}. 

If now instead of an isolated model Hamiltonian we study a parametric family of
Hamiltonians, the space of parameters naturally splits into regions with each of which 
is associated the set of constant Chern numbers for isolated bands \cite{IwaiAnnPhys}.   
The boundaries between different regions in the control parameter space correspond to 
a model Hamiltonian with degeneracy in eigenvalues \cite{VPVdp}. 
Eigen-line bundles are not defined in such situation
and the system of Chern numbers is a piece-wise constant function defined almost everywhere 
on the space of control parameters. The region where Chern numbers are defined and 
keep their values constant  is called an iso-Chern domain \cite{IwaiAnnPhys}. 
The boundaries of iso-Chern domains are responsible for the modification of the band structure. 
Thus, a modification formula attached to the crossing of the boundary may be referred to as 
a wall-crossing formula.  

A straightforward  approach to a characterization of qualitative structure of bands
within each iso-Chern domain consists naturally in calculating Chern numbers
of eigen-line bundles for a suitably chosen representative Hamiltonian from each iso-Chern domain
\cite{IwaiAnnPhys}. 
However,  this procedure gets much more complicated, if the size of the matrix Hamiltonian 
becomes larger. On this account, 
it is much interesting to study only modifications in Chern numbers in association with 
crossing (in the control parameter space) the boundary between iso-Chern domains.  
The degeneracy in eigenvalues, which occurs when the parameter values fall on the boundary 
of the iso-Chern domain, gives rise to a local singularity for the splitting of eigen-line bundles.  
A topological (more exactly homotopy \cite{IwaiToBe}) invariant, which we call a delta-Chern 
invariant or simply a delta-Chern, 
can be assigned to crossing each boundary of iso-Chern domains. 
It shows how Chern numbers of individual eigen-line bundles change 
(in other words, how the qualitative structure of bands changes) when passing 
from one iso-Chern domain into another by boundary crossing.  

In what follows, we demonstrate how this approach, which is based on ``delta-Chern" 
invariants  attached to boundary crossing,  works well in relatively simple
molecular examples in the presence of finite symmetry groups, which are important
ingredients of the formulation of a molecular problem. 

With the present article, authors hope to bring attention 
of theoretical physicists and mathematicians
working on general wall-crossing phenomena to (close in spirit) physical problem
of qualitative description of finite particle systems  which allows a detailed
comparison  of predictions of qualitative models with a rich available
quantitative experimental information and suggests 
a natural way to pass to more general mathematical models.

\section{An example of three state problem}
We take here as an example an effective rotational Hamiltonian
for three vibrational states of $F_2$ symmetry in the presence of the cubic group $O$ 
(see \cite{Iwai} for similar effective Hamiltonians for 
two-dimensional vibrational representations 
of the $O$ symmetry group).

In terms of tensor products
of rotational and vibrational irreducible (with respect to the $O$ symmetry group) 
tensor operators, 
this effective Hamiltonian can be written as  
\begin{eqnarray}\label{Eq1}
H = \left[V^{A_1}\otimes R^{A_1}\right]^{A_1} + 
\left[V^{E}\otimes R^{E}\right]^{A_1} 
+ \left[V^{F_1}\otimes R^{F_1}\right]^{A_1} 
+ \left[V^{F_2}\otimes R^{F_2}\right]^{A_1}, 
\end{eqnarray} 
where $V^{\Gamma}$ and $R^{\Gamma}$ are vibration and rotation tensor 
operators, respectively, transforming according to irreducible
representations $\Gamma=A_1, A_2, E, F_1, F_2$ of the symmetry group 
$O$. Only invariant tensor operators (of $A_1$ symmetry type)
are needed to be included in the Hamiltonian.

Taking only leading contributions (the operators of lowest 
degree in $\{J_x,J_y,J_z\}$ variables) for 
rotational tensor operators and neglecting scalar
$J^2$ dependence,  
we have the following explicit form of 
rotational contributions in the classical limit, 
\begin{eqnarray}\label{Eq3}
R^{A_1} &=& J_x^4+J_y^4+J_z^4; \\
R^{E}_1 &=& 2J_z^2-J_x^2-J_y^2; \quad R^{E}_2 = \sqrt3 (J_x^2-J_y^2);\\
 R^{F_1}_x &=& J_x; \quad   R^{F_1}_y = J_y; \quad  
                     R^{F_1}_z = J_z; \\
 R^{F_2}_x &=& J_yJ_z; \quad   R^{F_2}_y = J_zJ_x; \quad  
                     R^{F_2}_z = J_xJ_y. 
\end{eqnarray} 


Vibrational operators in their turn can be expressed,  
for a fundamental band of $F_2$ symmetry, as 
bilinear products of elementary annihilation and 
creation harmonic oscillator operators of $F_2$ symmetry type
with respect to the  symmetry group $O$ of the problem. 

Now we are in a position to give explicit form to \eqref{Eq1}. 
To make consistent contributions from rotational operators of different degrees
to a complete Hamiltonian and to keep the Hamiltonian reasonably simple,  
we may restrict ourselves to operators of at most second degree. 
This allows us to write different contributions of (\ref{Eq1})
in a matrix form with only two phenomenological parameters
$a,b$ which we consider below as control parameters of the model.
 For the sake of simplicity of notations, we replace  below $J_\alpha=\alpha$, 
where $\alpha=x,y,z$. 
Then, our Hamiltonian is expressed as 
\begin{eqnarray}\label{Eq4} \nonumber
&\left( \begin{array}{ccc} 0 &  i z & -i y \\
              -i z & 0 & i x   \\
               i y & -i x & 0 \end{array} \right) 
+ a \left( \begin{array}{ccc}
    y^2+z^2-2x^2 & 0 & 0  \\
    0& z^2+x^2-2y^2 & 0 \\
    0 & 0 & x^2+y^2-2z^2    \end{array}\right) \\
&+ b \left( \begin{array}{ccc} 0 &  xy & zx \\
              xy & 0 &  yz   \\
               zx & yz & 0 \end{array} \right) . 
\end{eqnarray}
As we can always choose $J_x^2+J_y^2+J_z^2 =const$
and renormalise the model Hamiltonian by imposing this constant to be
equal to 1, we may view the Hamiltonian (\ref{Eq4}) as defined on the unit sphere $S^2$; $x^2+y^2+z^2=1$. 
Our next step for the qualitative analysis of the Hamiltonian (\ref{Eq4})
consists in finding all values of control parameters corresponding
to degeneracy of eigenvalues of the Hamiltonian  (\ref{Eq4})
at some points of the classical phase space associated with rotational
variables. 

\section{Iso-Chern domains in the control parameter space}
The presence of symmetry group considerably simplifies  the search
for degeneracy points of eigenvalues, since the symmetry group acts also on the 
classical base space $S^2$ \cite{MichelCRAS,PhysRep1}.  
The action of the symmetry group stratifies the phase space $S^2$ into orbits with 
different stabilizers. 
Criterion for degeneracy in eigenvalues (zero of the discriminant) 
can be formulated independently of  strata. 
At the same time, the solution can be immediately found for each of zero-dimensional 
strata by restricting the Hamiltonian on each of strata.  

\begin{figure}
\begin{center}
\includegraphics[width=0.5\columnwidth]{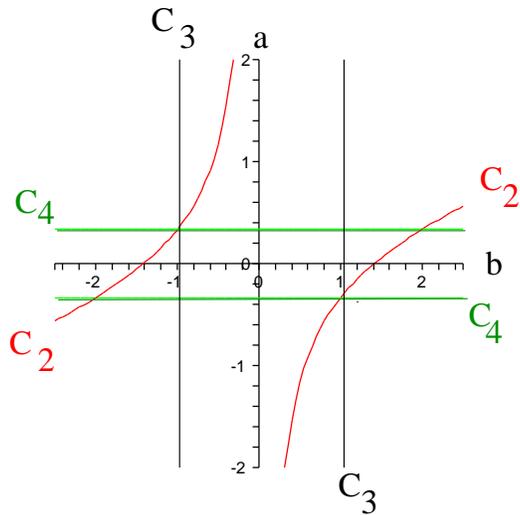} 
\end{center}
\caption{Degeneracy points (on $C_2,C_3,C_4$ orbits) in
the space of control parameters $(a,b)$ for the Hamiltonian 
\protect{(\ref{Eq4})}. 
\label{abPlotEFF} }
\end{figure}

From the discriminant of the characteristic equation for the Hamiltonian \eqref{Eq4},  
we find that the set of control parameters for which the Hamiltonian possesses degenerate 
eigenvalues is given by the curves or lines shown in Fig.~\ref{abPlotEFF}.  
For each point of the line $C_i$ in the space of control parameters, 
the Hamiltonian  \eqref{Eq4} possesses degenerate eigenvalues at a finite number of 
points of $S^2$, which form the orbit whose stabilizer is isomorphic to a cyclic group $C_i$. 
We call such an orbit a $C_i$ orbit. 
There are one $C_4$ orbit formed by six points which belong to three axes
passing through the centers of the opposite faces of the inscribed cube; 
one $C_3$ orbit formed by 8 points belonging to four axes passing through 
the four pairs of opposite vertices of the cube;  and one $C_2$
orbit formed by 12 points belonging to the six axes passing through the 
middle of the opposite edges of the cube.   
Control parameter values (for which  
the degeneracy points exist  on respective orbits) have simple expressions given by 
\begin{eqnarray} \label{ZeroStartDeg}
C_4\ : \ \ a=\pm\frac13; \quad C_3 \ : \ \ b=\pm1; \quad
C_2 \ : \ \ a = \frac{b^2-2}{3b}.
\end{eqnarray} 
Evaluating energy eigenvalues  at the degeneracy points,  one obtains 
\begin{eqnarray} \label{ZeroStartEnDEg}
&E_{C_4}(a,b) = (a-1,\ -2a, \ a+1); \quad
E_{C_3}(a,b)= \left( -1-\frac13 b, \ 1-\frac13 b, \frac23 b \right); \\ 
\label{ZeroStartEnDEg9}
&E_{C_2}(a,b)= \left(\frac{a-b-\sqrt{(3a+b)^2+16}}{4}, \ \frac{b-a}{2}, \ 
 \frac{a-b+\sqrt{(3a+b)^2+16}}{4} \right). 
\end{eqnarray} 
This tells us between which bands the degeneracy occurs. 

Intersection of two, $C_i$ and $C_j$,  lines corresponds to
control parameter values for which the Hamiltonian  \eqref{Eq4} possesses simultaneously
degeneracy points at both $C_i$ and $C_j$ orbits. The presence of points of 
simultaneous intersection of three lines is due to a non-generic feature 
of the model Hamiltonian whose entries are restricted within the terms up to second degree of
rotational contributions.  For these special control parameter values,  
the model Hamiltonian (\ref{Eq4}) possesses a  continuous set of degeneracy points on $S^2$. 
However, a small deformation of the Hamiltonian (\ref{Eq4}) by adding
higher order terms removes the triple intersection of degeneracy lines and puts the Hamiltonian 
in such a generic situation that a family of Hamiltonians possesses only generic (simple) intersection of 
degeneracy lines in the control parameter space.

Six lines $C_i$ shown in Figure \ref{abPlotEFF} split the plane of control parameters 
into 15 open simply-connected regions. Each region corresponds to one qualitative type of 
band structure formed by three energy bands. We characterize the qualitative type of 
the energy band by the Chern number for the eigen-line bundle associated with each of isolated
eigenvalues of the model Hamiltonian.  Since Chern numbers of eigen-line bundles
cannot change against a change in parameters without forming degeneracy points of eigenvalues, 
these regions are called iso-Chern domains.   

To characterize qualitatively the system of energy bands in each iso-Chern
domain, one needs to calculate Chern numbers for each of eigen-line bundle 
for a Hamiltonian at one particular set of values of control parameters. 
At the same time it is clear that at the boundary of iso-Chern domain
Chern numbers are not defined only for those bands which correspond 
to a couple of degenerate eigenvalues, as is observed from 
(\ref{ZeroStartEnDEg}) and (\ref{ZeroStartEnDEg9}).  
This fact allows us to simplify significantly
the study of a family of Hamiltonians near the boundary of iso-Chern
domain. Only two bands are to be taken into account. The third one
or in N-state problem all except two can be neglected. Moreover, as long  
as we are interested in qualitatively  describing a modification in Chern number, 
which should be an integer invariant, 
in crossing the boundary of an iso-Chern domain, small
perturbations caused by neglecting eigenvalues distant from 
the two degenerate ones in question cannot change the integer invariant.  
We call such an invariant a delta-Chern invariant or simply a delta-Chern.  
From the mathematical point of view, the delta-Chern is a homotopy invariant 
rather than simply a topological invariant \cite{IwaiToBe}.  It is important to note here that 
owing to the symmetry group action the degeneracy points appear simultaneously
on all points belonging to the same orbit of the symmetry group.  
Taking into account the fact that degeneracy points are generically isolated for a finite group,  
we can restrict the calculation of the delta-Chern invariant to a small neighborhood of 
the degeneracy point in the space of classical variables (small neighborhood of a point on the 
$S^2$ classical phase sphere) and to a small neighborhood of a degeneracy
point in the space of control parameters. This explains why we can 
simplify significantly the calculation of the delta-Chern invariant by
restricting to a local two-state Hamiltonian and even by making the 
linearization of the Hamiltonian near the degeneracy point.

\section{Local Hamiltonians near degeneracy point and associated 
delta-Chern invariant}
The importance of the analysis of local behavior of eigenvalues and eigenvectors
near the degeneracy point for the description of the qualitative phenomenon
of redistribution of energy levels between bands was noticed in initial
works on the redistribution phenomenon \cite{VPVdp,MolPhys88,FaurePRL}. The
systematic procedure for the calculation of Chern numbers for 
isolated bands in the absence of degeneracy points was suggested in
\cite{IwaiAnnPhys}.  It is performed by calculating all contributions
from so-called exceptional points which are associated with singularities 
of eigenvectors defined open-densely on $S^2$.  
Though the method to be applied in the case of the absence of 
degeneracy points cannot be applied if degeneracy appears, a modified 
approach can be made so as to be applicable in the case of the
presence of degeneracy points of eigenvalues. The application of the 
modified approach to delta-Chern calculation relies  on the
appropriate choice of the basis for the matrix Hamiltonian and 
of the basis of the tangent plane to the $S^2$ base space.
If such a set-up is well made, the modification of a system of 
exceptional points (at which eigenvectors are not defined) 
against a change in the control parameters may be observed in a small neighborhood  
of the degeneracy point in question, and hence the delta-Chern invariant can be  
calculated explicitly within the local approach. 

In general, if a symmetry group $G$ and its action on the classical phase space 
(the base space of the fiber bundle in the semi-quantum model) is known, 
the possible local symmetry is also known at points 
of the classical phase space. 
If the local symmetry (stabilizer) at a given point is a subgroup $G_i \subset G$, 
the number of point in the orbit is $[G]/[G_i]$, where $[G_i]$ is the order of group $G_i$. 
In the present case, $G$ is the $O$ group and $G_i$ is a cyclic group $C_i$.  
Taking into account the above-mentioned fact that the degeneracy points appear simultaneously
at all points of the orbit and the delta-Chern contributions from different
points of the same orbit are identical, we get immediately certain restrictions
on possible modifications of the band structure for a given symmetry group.

\begin{figure}
\begin{center}
\includegraphics[width=0.95\columnwidth]{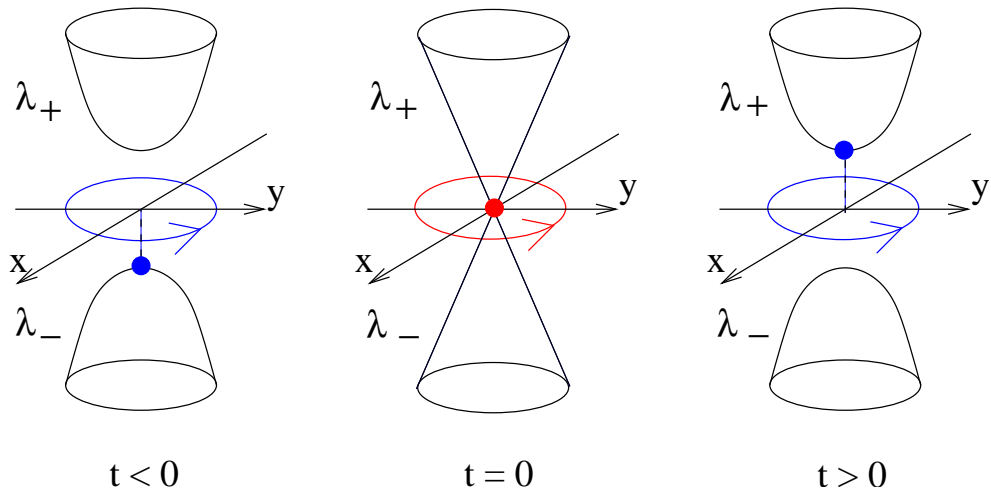} 
\caption{Schematic representation of the evolution of eigenvalues
of a local linearized model Hamiltonian in
a two-level approximation  along with
variation of a control parameter $t$ crossing the boundary
of the iso-Chern domain. Exceptional points (blue
points) in the chosen representation are shown for 
$\lambda_+$ and $\lambda_-$ components.  
  \label{DeltaChB}}
\end{center}
\end{figure}

\begin{figure}
\begin{center}
\includegraphics[width=0.4\columnwidth]{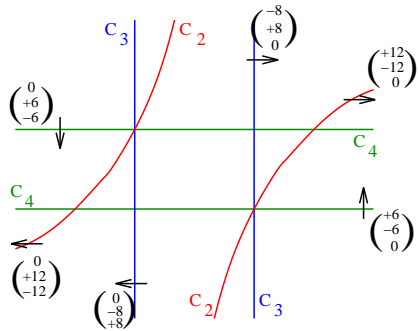} 
\end{center}
\caption{ Delta-Chern diagram for three state model \protect{(\ref{Eq4})} represented
in the space of control parameters $(a,b)$. Each degeneracy 
line (the boundary of the iso-Chern domain) is associated
with  a three component column giving delta-Chern for each of three bands 
and with an arrow indicating the direction of the path in the control parameter 
space associated with the shown modifications of Chern numbers. 
\label{CherDia3stDelta}}
\end{figure}

Figure  \ref{DeltaChB} illustrates a typical modification in exceptional points 
for a family of local Hamiltonians defined in a neighborhood 
of a degeneracy point. The details of the mathematical analysis through this typical 
modification will be given in \cite{IwaiToBe}, 
where particular attention is paid to the sign of the local
contribution to the delta-Chern. The global value of the delta-Chern follows directly
from local contribution by multiplying local contribution by the
order of the orbit of the degeneracy point. 
For our Hamiltonian \eqref{Eq4}, the delta-Cherns in crossing the boundaries 
are shown in Fig.~\ref{CherDia3stDelta}.

We are now in a position to make a qualitative description of band rearrangement. 
So far we have associated with each boundary the corresponding delta-Chern invariant 
(see Figure \ref{CherDia3stDelta}), it is 
sufficient for us to calculate the Chern numbers for just
one point in the control parameter space which can be chosen to
assure the simplest possible form of the effective Hamiltonian.

Combining this single point calculation of Chern numbers 
with all known delta-Chern invariants, we can easily reconstruct
the complete Chern diagram to give Chern indices for all
bands in all iso-Chern domains. Note that the sum of Chern numbers for 
isolated line bundles remains invariant when the control parameter
value passes a degeneracy point. In particular, the sum of the whole 
set of Chern numbers for the set of associated line bundles in
this paper is zero, and further, the symmetry restrictions do not allow the 
trivial vector bundle to decompose generically into a sum of trivial line
bundles.

\begin{figure}
\begin{center}
\includegraphics[width=0.5\columnwidth]{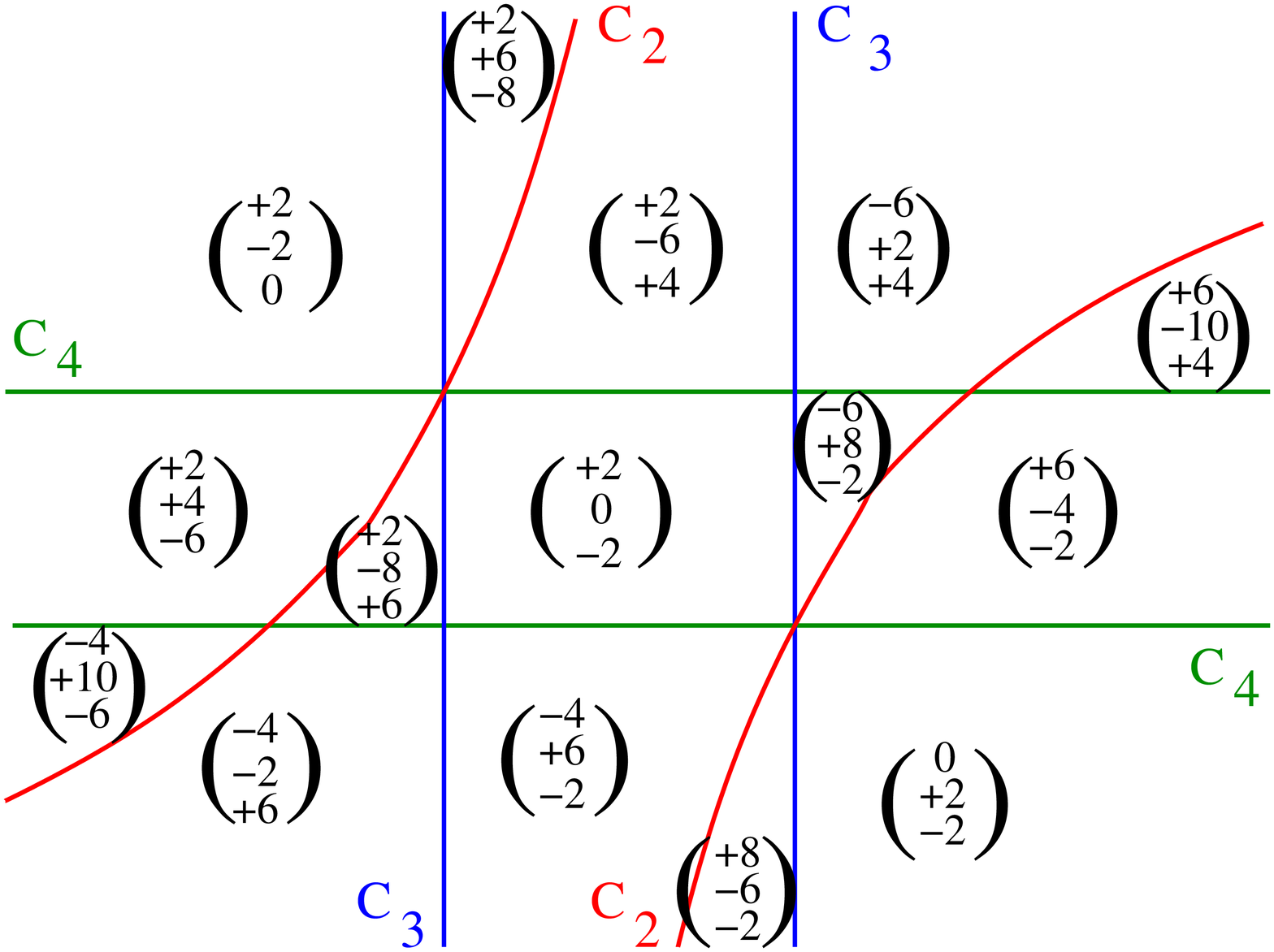} 
\end{center}
\caption{ Iso-Chern diagram for three state model. 
Vertical (blue) lines represent boundaries with degeneracy points at
$C_3$ positions. Horizontal (green) lines represent boundaries with 
degeneracy points at $C_4$ positions. Curved (red) lines represent boundaries with 
degeneracy points at $C_2$ positions.
Each open iso-Chern domain
is characterized by three Chern numbers associated with three
bands arranged according to their energy. Upper, middle and
lower numbers in each symbol give respectively 
Chern numbers for the band with higher, middle and lower energy. 
\label{ChernDia3st}}
\end{figure}

The Figure \ref{ChernDia3st} is obtained in this manner 
and gives a complete information on possible band rearrangements
under the variation in control parameters of the initial Hamiltonian
(\ref{Eq4}).

So far we have treated mainly a three-level Hamiltonian from the
view point of delta-Chern. The concept of delta-Chern is valid also
for $N$-level models.
Returning now to the example represented in Fig. \ref{FigEx3States}, we
illustrate qualitative description of band rearrangements occurring in
Fig. \ref{FigEx3States} using the correspondence between the number of
quantum states in bands and the associated Chern numbers as given in 
Table \ref{Table_N_Chern}  for low $J\sim8$ and high $J\sim30$ values.
Though the correspondence has not been given a rigorous proof, it
is supported by a number of examples.

\begin{figure}
\begin{center}
\includegraphics[width=0.29\columnwidth]{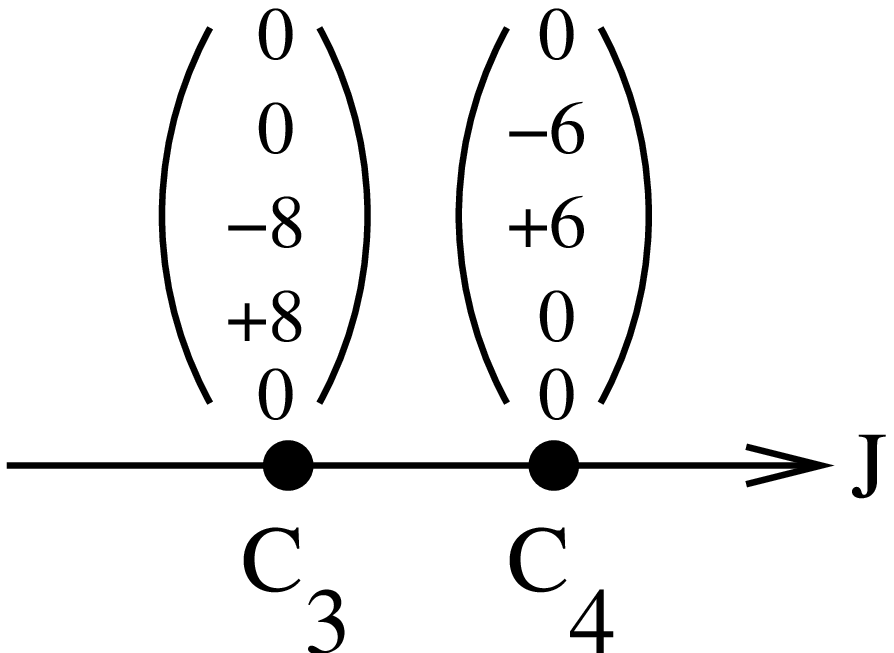} 
\hskip1cm
\includegraphics[width=0.4\columnwidth]{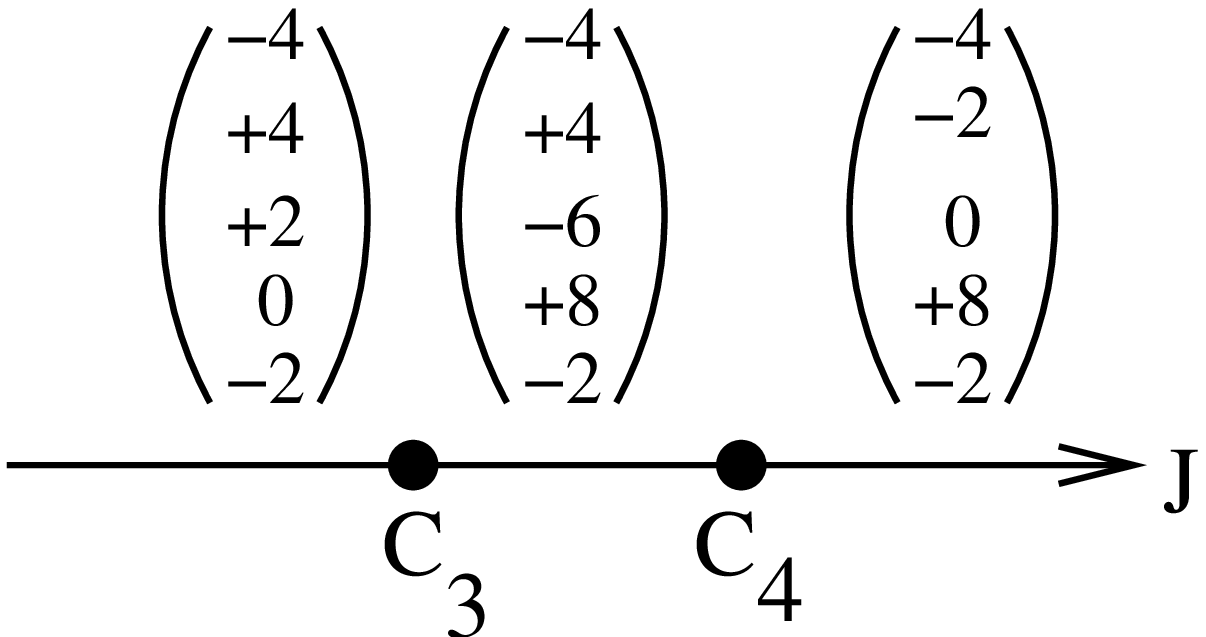}
\end{center}
\caption{Schematic representation of the evolution of
the band structure represented in Figure 1 as a function of
one control parameter, the rotational quantum number $J$. 
Left - delta Chern associated with two, $C_4$ and $C_3$,
orbits of degeneracy points. Right - Chern
numbers for isolated line bundles vs control parameter $J$.
\label{5Chern} }
\end{figure}

Without forming an explicit semi-quantum model, we assume that there is
an appropriate five-level model semi-quantum Hamiltonian and the delta-Chern
formula applies to this model. Then, from Fig. \ref{FigEx3States} it turns
out that the formation of at least 
two types ($C_3$ and $C_4$) of degeneracy points
is needed to pass from the low-$J$ to the high-$J$ band structure. The corresponding
systems of delta-Chern  together with  values of Chern numbers
are shown in Fig. \ref{5Chern}. They are consistent with 
the well-known decomposition
of doubly and triply degenerate bands for tetrahedral molecules at low $J$
values. 

\begin{table}
\caption{Energy bands and corresponding Chern numbers for example
shown in figure \protect{\ref{FigEx3States}}.  \label{Table_N_Chern} }
\vskip10pt
\begin{tabular}{l|cc|cc}
Band &  $J\sim 8$ & $J\sim 8$ &  $J\sim 30$ &  $J\sim 30$   \\
     & Numb. lev. & Chern numb. & Numb. lev. & Chern numb. \\ \hline
$\nu_2$ (upper) &  $2J-3$ & $-4$ & $2J-3$& $-4$ \\
 $\nu_2$ (lower) & $2J+5$ & $+4$ &  $2J-1$ &  $-2$  \\
$\nu_4$ (upper) & $2J+3$ & $+2$ &  $2J+1$ &  $0$ \\
$\nu_4$ (middle) & $2J+1$ & $0$ &   $2J+9$  & $+8$ \\
$\nu_4$ (lower) &   $2J-1$ & $-2$&  $2J-1$  & $-2$   \\ \hline
\end{tabular}
\end{table}

Another interesting and important conclusion from the viewpoint of applications is that 
possible rearrangements of the band structure can be formulated 
for problems with certain symmetry. 
We have already found a possible rearrangement in the name of delta-Chern in the case 
where the symmetry group is the $O$ group. 
Interesting implication in generic situation is the existence of selection rules for possible
values of Chern numbers for bands in the presence of symmetry, if we are
looking for possible modifications of several bands forming initially trivial
vector bundle with zero Chern numbers. Several examples of such selection
rules for Chern numbers of individual bands in the presence of symmetry
was formulated in \cite{Iwai}. The so-obtained selection rules for Chern
numbers reproduce similar selection rules for possible types of
vibrational components in molecular systems in the presence of symmetry
\cite{BrodSym}, which are based on reduction induction of representations
in elementary group theory. Possible relation with McKay correspondence
\cite{McKay1,McKay2} remains an interesting point to study.  

\section{Possible generalizations}
The above-studied particular example of model Hamiltonian 
(\ref{Eq4}) is rather a simple example in several factors.  It 
should be modified so as to allow application to a larger class of
molecular examples.

Rather serious restriction is the dimension of the space of classical variables.
Physical examples where ``slow" variables are rotational variables
describing molecular rotation  lead to a classical phase space of dimension  
two (one degree of freedom).  A number of other molecular
effective Hamiltonians describing band structure are possible, where the 
number of degrees of freedom for slow variables is larger than one. 
Rather general model describes, for example, vibrational polyads 
(formed by excited states of a multidimensional
isotropic oscillator) for a molecule possessing several close 
lying electronic states \cite{FaureCP2}. 
Depending on the number $N$ of vibrational degrees of freedom, the 
internal structure of vibrational polyads is described in the classical limit
by a model with compact phase space diffeomorphic 
to $\mathbb{C}P^{N-1}$ \cite{CP2Zhil}. 
If, as it often occurs in molecular problems, the frequencies of
vibrations are in resonance relation $n_1\nu_1 = n_2\nu_2 = ...$,  
the corresponding classical phase space for slow vibrational motion 
is a weighted projective space.  Although molecular systems which can be described
within such direction of generalization certainly exist, the preliminary
construction of mathematical formalism adapted to qualitative study of such
molecular models should be done in order to understand how one can work
with such more complex systems within the  qualitative approach.

\end{document}